\documentclass[11pt]{article}

\usepackage{authblk}
\usepackage[font=small]{caption}

\let\citeleft=( 
\let\citeright=)

\usepackage[english]{babel}
\usepackage[letterpaper,margin=1in]{geometry}
\usepackage[utf8x]{inputenc}
\usepackage[T1]{fontenc}
\usepackage{bm}
\usepackage{amssymb}
\usepackage{xr}

\usepackage{amsmath}

\usepackage[hiresbb]{graphicx}
\usepackage{amsfonts}
\usepackage{algorithmic}
\usepackage{textcomp}
\usepackage{booktabs}

\usepackage{ifthen}
\usepackage[colorlinks=false,linkcolor=.,hypertexnames=false,hidelinks,breaklinks]{hyperref}
\usepackage{colortbl}
\usepackage[caption=off]{subfig}
\usepackage[labelfont=bf]{caption}
\usepackage{authblk}
\usepackage[numbers,comma,sort&compress]{natbib}
\usepackage{cleveref}
\usepackage{mathtools}
\newtagform{brackets}{[}{]}
\usetagform{brackets}
\usepackage{empheq}

\setcitestyle{round}

\DeclareMathOperator*{\argmin}{argmin}
\crefname{SuppInfoSection}{Supporting Information Section}{Supporting Information Section} 
\crefname{NormalSection}{Section}{Section} 

\begin{document}

\newcommand{\eqname}[1]{\tag*{#1}}

\newcommand{\bv}[1]{\mathbf{#1}}
\newcommand{\Grad}{\mbox{\boldmath $\nabla$}}
\newcommand{\Div}{\Grad\!\cdot}
\newcommand{\Lap}{\mbox{\boldmath $\Delta$}}
\newcommand{\Curl}{\Grad\!\times}
\newcommand{\delsq}{\nabla^2}
\newcommand{\mean}[1]{\left\langle{#1}\right\rangle}
\newcommand{\infd}{\textrm{d}}
\newcommand{\diff}[2]{\frac{\infd #1}{\infd #2}}
\newcommand{\difftwo}[2]{\frac{\infd^2 {#1}}{\infd {#2}^2}}
\newcommand{\pdiff}[2]{\frac{\partial #1}{\partial #2}}
\newcommand{\pdiffcon}[3]{\left(\frac{\partial #1}{\partial #2}\right)_{#3}}
\newcommand{\pdifftwo}[2]{\frac{\partial^2 {#1}}{\partial {#2}^2}}
\newcommand{\pdifftwomix}[3]{\frac{\partial^2 {#1}}{\partial {#2}\,\partial {#3}}}
\newcommand{\vel}{\bv{v}}
\newcommand{\expu}[1]{\textrm{e}^{#1}}
\newcommand{\ehat}{\bv{e}}
\newcommand{\intomega}{ \mathop{}\!\int\limits_{\Omega}}
\newcommand{\intelement}{ \mathop{}\!\int\limits_{e_k}}
\newcommand{\intelementgen}{ \mathop{}\!\int\limits_{e_k^{xy}}}
\newcommand{\intelementref}{ \mathop{}\!\int\limits_{e_k^{\xi \eta}}}
\newcommand{\be}{\begin{equation}}
\newcommand{\ee}{\end{equation}}
\newcommand{\baa}{\begin{alignat}{2}}
\newcommand{\eaa}{\end{alignat}}
\newcommand\smathcal[1]{
  \mathchoice
    {{\scriptstyle\mathcal{#1}}}{{\scriptstyle\mathcal{#1}}}{{\scriptscriptstyle\mathcal{#1}}}{\scalebox{.7}{$\scriptscriptstyle\mathcal{#1}$}}}
\newcommand\sbmathcal[1]{
  \mathchoice
    {{\scriptstyle{\bm{\mathcal{#1}}}}}{{\scriptstyle{\bm{\mathcal{#1}}}}}{{\scriptscriptstyle{\bm{\mathcal{#1}}}}}{\scalebox{.7}{$\scriptscriptstyle{\bm{\mathcal{#1}}}$}}}

\newcommand{\note}[2]{ {\bf #1: }#2} 

\renewcommand{\sectionautorefname}{Section}
\renewcommand{\figureautorefname}{Figure}\let\subsectionautorefname\sectionautorefname
\let\subsubsectionautorefname\sectionautorefname
\definecolor{lightgray}{rgb}{0.925,0.925,0.925}

\def\equationautorefname#1#2\null{Eq.#1[#2\null]}

\newcommand*{\vertbar}{\rule[-1ex]{0.5pt}{2.5ex}}
\newcommand*{\horzbar}{\rule[.5ex]{2.5ex}{0.5pt}}
\newcommand{\refappendix}[1]{\hyperref[#1]{Appendix~\ref*{#1}}}
\newcommand{\refsupinf}[1]{\hyperref[#1]{Supporting Information \ref*{#1}}}
\newcommand{\refundef}{{\color{red}[x]}}

\crefalias{subsection}{NormalSection}
\crefalias{section}{NormalSection}

\newcommand{\beginsupplement}{\setcounter{figure}{0}
    \setcounter{equation}{0}
    \renewcommand{\theequation}{S\arabic{equation}}
\crefalias{section}{SuppInfoSection}
\crefalias{subsection}{SuppInfoSection}
    \setcounter{section}{0}
}

\newcommand{\setsuppvideo}{\renewcommand*{\figurename}{\hspace{-0.3em}} \renewcommand*{\thefigure}{Supporting Information Video S\arabic{figure}}}

\newcommand{\setsuppfig}{\renewcommand*{\figurename}{\hspace{-0.3em}}\renewcommand*{\thefigure}{Supporting Information Figure S\arabic{figure}}}

\renewcommand*{\figurename}{\hspace{-0.5em}}
\renewcommand*{\thefigure}{Figure \arabic{figure}}

\renewcommand\Authands{ and }

\newcommand{\colorred}[1]{{\color{red}{#1}}}

\title{\vspace{-2cm}Non-rigid 3D motion estimation at high temporal resolution from prospectively undersampled k-space data using low-rank MR-MOTUS}

\author[1,2]{Niek R.F. Huttinga}
\author[1,2]{Tom Bruijnen}
\author[1,2]{Cornelis A.T. van den Berg}
\author[1,2]{Alessandro Sbrizzi}

\affil[1]{\small Department of Radiotherapy, Division of Imaging \& Oncology, University Medical Center Utrecht, The Netherlands}
\affil[2]{\small Computational Imaging Group for MR Diagnostics \& Therapy, Center for Image Sciences, University Medical Center Utrecht, The Netherlands}
\maketitle

\vfill
\noindent
\textit{Running head:} Non-rigid 3D motion estimation using low-rank MR-MOTUS

\noindent
\textit{Address correspondence to:} \\
  University Medical Center, Utrecht, Heidelberglaan 100, 3584 CX, Utrecht, The Netherlands \\
  n.r.f.huttinga@umcutrecht.nl

\noindent
This work was supported in part by the Dutch Research Council (NWO) under Grant 15115.

\noindent
Approximate word count: 247 (Abstract), 4992 (Main body)\\

\noindent
Submitted to \textit{Magnetic Resonance in Medicine} as a Full Paper.

\clearpage

\section*{Abstract}

\noindent
\textbf{Purpose}: With the recent introduction of the MR-LINAC, an MR-scanner combined with a radiotherapy LINAC, MR-based motion estimation has become of increasing interest to (retrospectively) characterize tumor and organs-at-risk motion during radiotherapy. To this extent, we introduce low-rank MR-MOTUS, a framework to retrospectively reconstruct time-resolved non-rigid 3D+t motion-fields from a single low-resolution reference image and prospectively undersampled k-space data acquired during motion.

\noindent
\textbf{Theory}: Low-rank MR-MOTUS exploits spatio-temporal correlations in internal body motion with a low-rank motion model, and inverts a signal model that relates motion-fields directly to a reference image and k-space data. The low-rank model reduces the degrees-of-freedom, memory consumption and reconstruction times by assuming a factorization of space-time motion-fields in spatial and temporal components.

\noindent
\textbf{Methods}: Low-rank MR-MOTUS was employed to estimate motion in 2D/3D abdominothoracic scans and 3D head scans. Data were acquired using golden-ratio radial readouts. Reconstructed 2D and 3D respiratory motion-fields were respectively validated against time-resolved and respiratory-resolved image reconstructions, and the head motion against static image reconstructions from fully-sampled data acquired right before and right after the motion.

\noindent
\textbf{Results}: Results show that 2D+t respiratory motion can be estimated retrospectively at 40.8 motion-fields-per-second, 3D+t respiratory motion at 7.6 motion-fields-per-second and 3D+t head-neck motion at 9.3 motion-fields-per-second. The validations show good consistency with image reconstructions.

\noindent
\textbf{Conclusion}: The proposed framework can estimate time-resolved non-rigid 3D motion-fields, which allows to characterize drifts and intra and inter-cycle patterns in breathing motion during radiotherapy, and could form the basis for real-time MR-guided radiotherapy.
  
\noindent
\textbf{Keywords}: Motion estimation, Model-based reconstruction, MR-guided radiotherapy, MR-LINAC

\clearpage


\section{Introduction}
Uncertainty in tumor and organs-at-risk locations due to unknown respiratory-induced organ motion diminishes the efficacy of radiotherapy in the abdomen and thorax in two ways. Firstly, tumors are irradiated with larger treatment margins, which results in increased radiation dose and toxicity to healthy tissue. Secondly, it prevents an accurate (retrospective) estimation of the actual dose accumulated in the targeted tumor and healthy surrounding tissue during the treatment. 

Recently, the MR-LINAC was introduced as the combination of an MR-scanner and a linear accelerator (LINAC) in a single device \citep{lagendijk2008mri,mutic2014viewray,Keall2014,raaymakers2009integrating}, which has the potential to address both points above. Achieving this goal, however, poses the following technical challenge: real-time reconstructions at 5 Hz \citep{keall2006management,Murphy2002} of internal body motion during the treatments. A fundamental step towards real-time reconstructions is the retrospective estimation of time-resolved motion-fields. Additionally, these retrospectively reconstructed motion-fields are valuable for the calculation of accumulated dose and can be taken into account for more accurate radiation planning of subsequent treatments. To this extent, we focus on the retrospective reconstruction of time-resolved 3D+t respiratory motion with a temporal resolution of 5 motion-fields-per-second. We envision that this framework could eventually be adapted to prospective real-time reconstructions \citep{huttinga2020realtime}.

In MR-guided radiotherapy, tumor and organs-at-risk motion is typically estimated from cine-MR-images followed by image registration. For time-resolved motion estimation, these cine-MR-images would thus require sufficient temporal resolution and spatial coverage to resolve the targeted motion. This is in general achievable in 2D, and also in 3D for slowly moving targets such as pelvic tumors \citep{de_Muinck_Keizer_2019}. However, in 3D it is more challenging for faster moving targets like lung tumors, that require at least 5 motion-fields-per-second \citep{keall2006management,Murphy2002}. 

Several strategies have previously been proposed to extract tumor and organ-at-risk motion from MR-images, three of which will be reviewed below. With the first strategy, average respiratory motion is estimated from a respiratory-resolved 3D+t MRI. This approach retrospectively sorts image slices or k-space readouts in 3D acquisitions according to their respective respiratory phases, extracted using a respiratory motion surrogate (e.g. pneumatic belt, self-navigation signal or navigator). Examples include the works in \citep{breuer2018stable,deng2016four,han2017respiratory,cai2011four,feng2016xd} (see \citep{stemkens2018nuts} for a more complete overview). Although the retrospective sorting in these methods allows for efficient use of all acquired data, it makes strong assumptions on the periodicity of respiratory motion and characterizes only average 3D+t breathing motion. Although this is useful to reduce treatment margins, it may not be sufficient for accurate accumulation of the delivered dose.
 
A different strategy uses multi-slice/orthogonal 2D+t cine-MRI for 3D+t motion estimation \citep{Paganelli2018,bjerre2013three,tryggestad20134d,brix2014three,stemkens2016image,seregni2016motion,mcclelland2013respiratory,mcclelland2017generalized}. The reduction in the spatial dimension allows for higher temporal resolution, and is combined with a model that links the lower-dimensional image data to 3D motion-fields. This strategy assumes, however, that a good fit on lower-dimensional images implies a good fit in the full 3D domain. Although this is reasonable for small volumes, since slices cover a large fraction of the volume in such a case, it may be less valid for larger volumes which may be required for dose accumulation.

The third strategy does not rely on sorting, but reconstructs images from highly undersampled k-space data. Even with parallel imaging \citep{pruessmann1999sense,griswold2002generalized}, this typically eventually results in lower SNR, lower spatial resolution, and/or undersampling artifacts. Nevertheless, it has been shown that motion-fields can be estimated from these images with sufficient accuracy  \citep{glitzner2015line,stemkens2013optical,roujol2011automatic,yuan2019fast}. Additionally, iterative reconstructions based on compressed sensing \citep{lustig2008compressed} have been proposed to exploit the spatio-temporal sparsity of images. However, for the intended application the reported temporal resolution was too low \citep{yuan2019fast,ong2019extreme, king2012thoracic}, or the FOV was too small \citep{fu2017high,burdumy2017one}.

Following a different approach, we have previously introduced MR-MOTUS \citep{huttinga2020mr} (Model-based Reconstruction of MOTion from Undersampled Signal), a new framework that allows to reconstruct whole-body non-rigid 3D motion-fields directly from k-space data. The key ingredient of MR-MOTUS is a signal model that explicitly relates dynamic k-space data to the combination of a static reference image and dynamic motion-fields. Assuming a reference image is available, and data is acquired in steady-state, motion-fields can be reconstructed directly from k-space data by solving the corresponding non-linear inverse problem. Since motion-fields are spatially correlated and therefore compressible, few data are required for the reconstructions. 

The possibility to reconstruct motion from few k-space data makes MR-MOTUS a natural candidate for time-resolved 3D+t motion estimation, which is not directly restricted to the achievable temporal resolution in MR-images. Our work presented in \citep{huttinga2020mr}, however, represents a proof-of-concept, and demonstrates MR-MOTUS in an experimental setting. Four points of improvement should be addressed for the extension of MR-MOTUS to time-resolved 3D+t motion estimation:

\begin{enumerate}
\item Only spatial correlation in motion-fields was exploited, and a single static motion-field was reconstructed for each single snapshot of k-space data. Additionally exploiting temporal correlation, and jointly reconstructing the 3D+t motion-field series at once, could improve the reconstruction quality and lower requirements of computing time and memory. 
\item Only the body coil was used for data acquisition to obtain homogeneous coil sensitivity. This did not represent a practical setting and multi-coil acquisitions would be favorable in terms of SNR.
\item The required reference image was obtained from a separate MR-scan during breath-hold. Ideally, no breath-holds are required and reconstructions can be performed on data acquired in free-breathing conditions.
\item 3D motion-fields were previously reconstructed from retrospectively undersampled Cartesian k-space data, while the motion estimation application requires prospectively undersampled acquisitions with an efficient non-Cartesian trajectory.
\end{enumerate}
In this work we address the aforementioned points of improvement and extend the framework to experiments in a realistic setting, in which reference image and time-resolved 3D+t motion-fields can be reconstructed from multi-coil, free-breathing, prospectively undersampled non-Cartesian 3D k-space data. The reconstruction of time-resolved motion-fields yields a large number of unknowns in 3D, which are needed to represent 3D motion-fields over a large number of timepoints (>100). We propose to use a spatio-temporal low-rank motion model, which yields a compressed representation of motion-fields in space and time. Several works have previously proposed low-rank motion models for motion estimation \citep{zhang2007patient, stemkens2016image, li2011pca, mishra2014initial,cai20153d,Low2005}, and the analyses in \citep{zhang2007patient, stemkens2016image, li2011pca} suggest that a rank-2 motion model can accurately describe respiratory motion. Consequently, the low-rank motion model can reduce the number of unknowns by two orders of magnitude, thereby introducing a regularization in both space and time and significantly reducing memory consumption and reconstruction times for 3D+t reconstructions. We will refer to the extended framework as low-rank MR-MOTUS.

We demonstrate and validate low-rank MR-MOTUS in a total of 6 {\it in-vivo} experiments on 2 healthy subjects and several moving anatomies. 2D/3D whole-body respiratory motion is included in view of the MR-guided radiotherapy application, and 3D head-and-neck motion is included for additional validation and as a demonstration to handle different types of motion. The 2D respiratory motion reconstruction is validated against 2D time-resolved compressed sensing, the 3D respiratory motion reconstruction against respiratory-resolved 3D image reconstruction, and the 3D head-and-neck motion against 3D static images acquired right before and right after the motion.

\section{Theory}
\label{section:theory}

\subsection{Background MR-MOTUS}
\label{section:theory_background}
We assume a general $d$-dimensional setting, with targeted case $d=3$, and we follow the convention that bold-faced characters denote vectorizations. We define $\bv{x}_0 \mapsto \bv{x}_t$ as the mappings from coordinates $\bv{x}_0\in\mathbb{R}^d$ in a reference image to new locations $\bv{x}_t\in\mathbb{R}^d$ at time $t$. The mappings are characterized by the motion-fields $\bv{d}_t$ through $\bv{x}_t = \bv{x}_0 + \bv{d}_t(\bv{x}_0)$. This will be written in concatenated vector-form as
\begin{equation}
    \bv{X}_t = \bv{X}_0 + \bv{D}_t,   
\end{equation}
where $\bv{X}_t,\bv{X}_0,\bv{D}_t \in \mathbb{R}^{N d \times 1}$ denote the vertical concatenations over $N$ spatial points in a $d$-dimensional setup.
The MR-MOTUS forward model \citep{huttinga2020mr} explicitly relates the motion-fields $\bv{D}_t$ and a static reference image $\bv{q}_0\in\mathbb{C}^N$ to dynamic, single-channel (and possibly non-Cartesian) k-space measurements $\bv{s}_t\in\mathbb{C}^{N_k}$: 
\begin{equation}
\bv{s}_t = \bv{F}(\bv{D}_t |\bv{q}_0) + \bm{\epsilon}_t.
\label{eq:signalmodel}
\end{equation}
Here $\bm{\epsilon}_t\in\mathbb{C}^{N_k}$ is the complex noise vector and $\bv{F}:\mathbb{R}^{Nd} \mapsto \mathbb{C}^{N_k} $ is the vectorization of the forward operator defined as
\begin{equation}
\label{eq:forwardmodel}
F(\bv{d}_t)[\bv{k}] = \int_\Omega q_0(\bv{x}_0) e^{-i 2 \pi \bv{k} \cdot \left[ \bv{x}_0 + \bv{d}_t(\bv{x}_0) \right]} \ \infd \bv{x}_0,
\end{equation}
where $\bv{k}\in\mathbb{R}^d$ denotes the k-space coordinate. By fitting the non-linear signal model in \autoref{eq:forwardmodel} to acquired k-space data, motion-fields can be reconstructed directly from k-space measurements. 

\subsection{Reconstruction problem formulation for space-time reconstructions}
\label{section:reconstructionproblemformulation}
In this work we follow \citep{huttinga2019prospective} and formulate the reconstruction problem for space-time motion-field $\bv{D}$ as follows: 
\begin{equation}
\label{eq:inverseproblemspatiotemporal}
\min_\bv{D} \ \sum_{t=1}^M \left\lVert \bv{F}( \bv{D}_t ) - \bv{s}_t \right\rVert_2^2 + \lambda_R\mathcal{R}(\bv{D}).
\end{equation}
Here $\mathcal{R}(\bv{D})>0$ is a regularization functional, with corresponding parameter $\lambda_R >0$, which models a-priori assumptions in order to exploit correlations in both space and time.

\subsubsection{Parameterization with a low-rank space-time motion model}
\label{section:lowrankmodel}

A straightforward parameterization of $\bv{D}$ considers one motion-field per dynamic, i.e. $\bv{D}=[\bv{D}_1,\dots,\bv{D}_M]\in\mathbb{R}^{Nd\times M}$. This is, however, impractical from a computational point-of-view, since the number of parameters scales with the number of dynamics: $|\bv{D}|=NMd \sim M$. For a typical scenario, $N\sim 10^6, M\sim 10^2$ and $d=3$, in which case 
\begin{equation}\label{eq:paramscaling}|\bv{D}| \sim 10^8.\end{equation}
Hence, this parameterization is inconvenient since it results in high memory consumption and long reconstruction times.

Instead, we propose a parameterization with a low-rank motion model, to simultaneously reduce the number of parameters for the reconstruction and introduce a natural regularization in both space and time. The model enforces the following factorization of the motion-fields that separates the spatial and temporal contributions:
\begin{alignat}{2}\bv{D} &= \left(\begin{array}{ccc} \vertbar & & \vertbar \\ \bv{D}_1 & \dots & \bv{D}_M \\ \vertbar & & \vertbar \\  \end{array}\right)
&= \left(\begin{array}{ccc} \vertbar & & \vertbar \\ \bm{\Phi}^1 & \dots & \bm{\Phi}^R \\ \vertbar & & \vertbar \\  \end{array}\right)  \left(\begin{array}{ccc} \horzbar & \bm{\Psi}^1 &   \horzbar \\ & \vdots & \\ \horzbar & \bm{\Psi}^R & \horzbar \end{array}\right)=\bm{\Phi}\bm{\Psi}^T.\label{eq:lowrankmotionmodel}
\end{alignat}
Here $R$ denotes the number of components of the model; $\bm{\Phi} \in\mathbb{R}^{Nd \times R}$ denotes the matrix with spatial components, and $\bm{\Psi}\in\mathbb{R}^{M \times R}$ denotes the matrix with temporal components. The model \eqref{eq:lowrankmotionmodel} will be referred to as the low-rank model, since $\text{rank}(\bv{D})\le R$. The upper limit is achieved for $R$ linearly independent components. A similar explicit low-rank factorization was recently proposed in the context of image reconstructions in \citep{ong2019extreme}, with the same motivations as mentioned above.

The number of parameters in the low-rank model is $|\bv{D}|=|\bm{\Phi}|+|\bm{\Psi}|=(Nd + M)R$. Analyses in the works \citep{zhang2007patient,stemkens2016image,li2011pca} suggest that $R=2$ is sufficient to accurately model respiratory motion. Hence, for the typical scenario considered above ($N\sim10^6,M\sim10^2,d=3$), this implies
\begin{equation}
    |\bv{D}|\sim 10^6,
\end{equation}
which is two orders of magnitude lower than \autoref{eq:paramscaling} and makes the reconstructions more convenient in practice.

We follow a standard approach in non-rigid medical image registration \citep{rueckert1999nonrigid} and represent both the spatial components $\bm{\Phi}$ and the temporal components $\bm{\Psi}$ of the motion-fields in cubic B-spline bases. This results in representation coefficients $\bm{\alpha},\bm{\beta}$ for respectively $\bm{\Phi}$ and $\bm{\Psi}$.

\subsubsection{Regularization functional}
The motion-field reconstruction problem in \autoref{eq:inverseproblemspatiotemporal} is typically ill-posed, and requires incorporation of a-priori knowledge of the motion-fields in terms of additional regularization terms. Since organs such as the liver, spleen and kidney consist of liquid filled tissue structures, they can be assumed incompressible and thus volume-preserving under motion \citep{zachiu2018anatomically}. This assumption can be incorporated into the reconstruction problem with the following regularization, based on the determinant of the Jacobian of the transformation corresponding to the motion-fields \citep{Rohlfing2003}:
\begin{equation}\label{eq:regularizationdetjac} \mathcal{R}(\bv{D}):= \sum_{t=1}^M \lVert \bv{W} ( \mathcal{J}(\bv{D}_t) - \bv{1} )\rVert_2^2.  \end{equation}
Here $\mathcal{J}(\cdot)$ computes the determinant of the Jacobian, and $\bv{W}$ is a diagonal matrix with weights per voxel. The weights are added to exclude regions where the regularization is less realistic, e.g. in the lungs. As weights we have taken the magnitude of the reference image, scaled to unit norm. For the implementation of this regularization term we follow the approach in \citep{Rohlfing2003}, and compute spatial derivatives analytically using the spline parameterization of the motion-fields.

\subsubsection{Final reconstruction problem formulation}
Substituting the spline representation, low-rank model \eqref{eq:lowrankmotionmodel} and regularization \eqref{eq:regularizationdetjac} into the objective function \eqref{eq:inverseproblemspatiotemporal} results in the following minimization problem to reconstruct space-time motion-fields:
\begin{equation}
\label{eq:inverseproblemspatiotemporal_compressed}
\{\bm{\alpha}^\dagger,\bm{\beta}^\dagger\} = \argmin_{\substack{ \bm{\Phi}\bm{\Psi}^T}=[\bv{D}_1,\dots,\bv{D}_M] } \ \sum_{t=1}^M \left\lVert \bv{F}( \bv{D}_t ) - \bv{s}_t \right\rVert_2^2 +  \lambda_R \sum_{t=1}^M \lVert \bv{W}(\mathcal{J}(\bv{D}_t) - \bv{1})\rVert_2^2,
\end{equation}
where $\lambda_R\in\mathbb{R}^+$ is the regularization parameter that balances the terms. Note that no temporal regularization is added, since the low-rank model already acts as a strong regularization in both space and time.

\section{Methods}

\noindent{\bfseries Experiments overview}\\
The following data were acquired in three different experiments per volunteer for two volunteers: 
\begin{enumerate}
    \item 2D+t abdominothoracic data;
    \item 3D+t abdominothoracic data;
    \item 3D+t head-and-neck data.
\end{enumerate} The 2D+t abdominothoracic data allows for a validation against time-resolved image reconstruction at a high temporal resolution. The 3D+t abdominothoracic data is the targeted case for the application in MR-guided radiotherapy. The 3D+t head-and-neck data is included as a demonstration to handle different types of motion, and for additional validation. All reconstructions are analyzed by comparison with image reconstructions on the same data. The Jacobian determinant of the transformation corresponding to the motion-fields is analyzed as an additional sanity check: $\bv{x}_0 \mapsto \bv{x}_0 + \bv{d}_t(\bv{x}_0)$. More details regarding the experiments is provided below, organized per subsection. 

\noindent{\bfseries Data acquisition}\\
All data were acquired on a 1.5T MRI scanner (Ingenia, Philips Healthcare, Best, the Netherlands) using a steady-state spoiled gradient echo sequence (SPGR) with anterior and posterior receive arrays. As readouts we employed golden-angle radial for 2D \citep{winkelmann2006optimal}, and golden-mean kooshball radial for 3D \citep{chan2009temporal}. The volunteers provided written informed consent prior to the scans, and all scans were approved by the institutional review board of the University Medical Center Utrecht and carried out in accordance with the relevant guidelines and regulations. See \autoref{tabel:experimentdetails} for other relevant acquisition parameters.  

\begin{figure}[h]
\centering
    \includegraphics[scale=0.155]{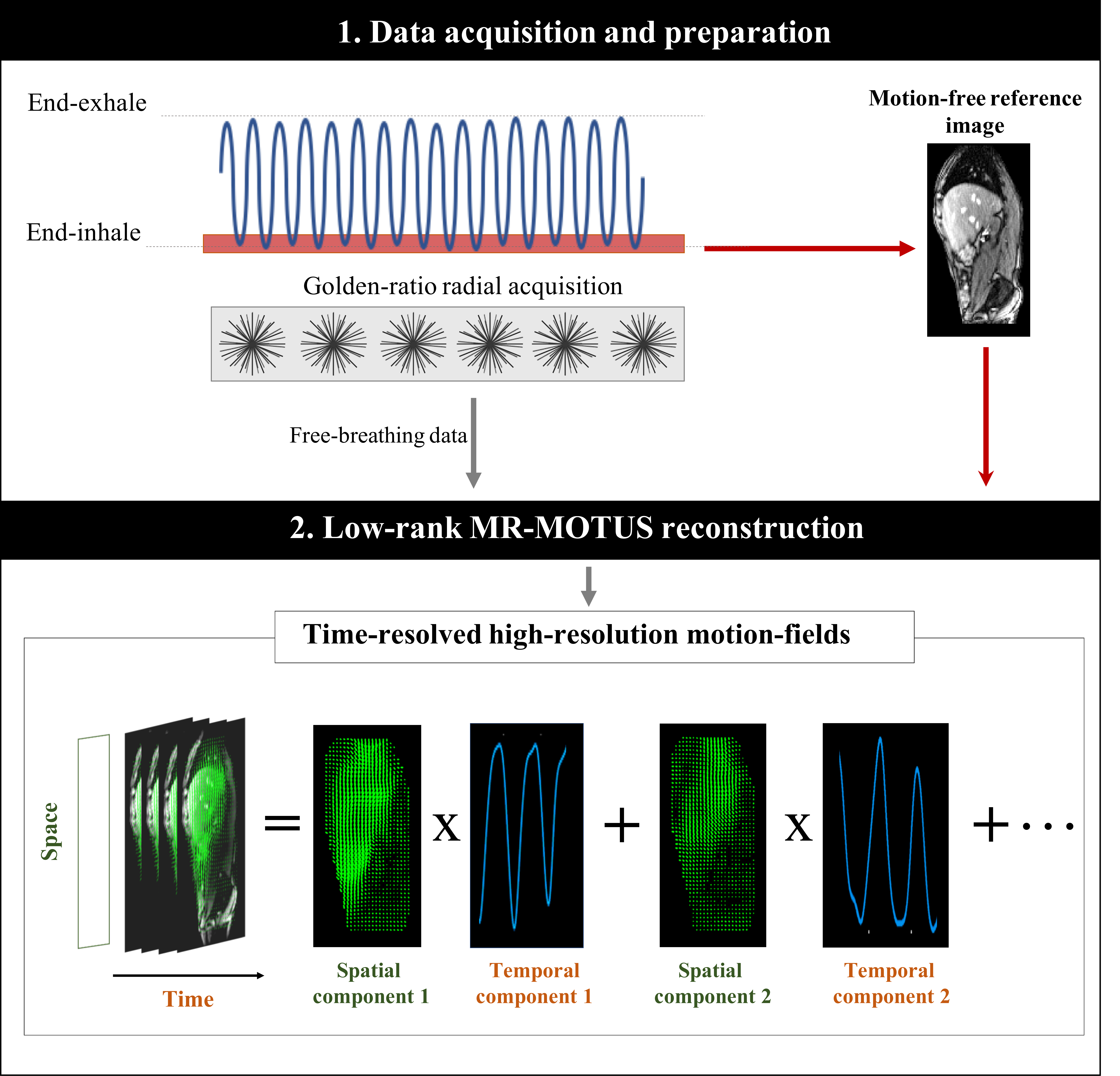}
    \caption{Overview of the low-rank MR-MOTUS framework. First, data is acquired during free-breathing with a golden-ratio radial trajectory (2D: golden-angle radial \cite{winkelmann2006optimal}, 3D: golden-mean radial kooshball \citep{chan2009temporal}). Then, DC-based phase-binning is performed on end-inhale to reconstruct a motion-free reference image. Finally, the reference image and free-breathing data are fed into the low-rank MR-MOTUS reconstructions, resulting in time-resolved 3D motion-fields. The motion-fields are reconstructed with an explicit constraint on the maximum rank. That is, as a sum of component motion-fields with each a different temporal behavior. The number of such components is pre-determined. }
    \label{figure:lowrankmodeloverview}
\end{figure}

\noindent{\bfseries Reconstruction details}\\
We followed the approach outlined in \cref{section:reconstructionproblemformulation}, and reconstructed motion-fields from multi-coil k-space data acquired during motion by solving the minimization problem \eqref{eq:inverseproblemspatiotemporal_compressed}. The low-rank MR-MOTUS workflow is schematically summarized in \ref{figure:lowrankmodeloverview}.
The reconstruction problem \eqref{eq:inverseproblemspatiotemporal_compressed} was solved with L-BFGS \citep{liu1989limited}, using the MATLAB implementation from \citep{beckerlbfgsb2019}. The L-BFGS memory parameter was set to 20 and sampling density compensation \citep{pipe1999sampling,pruessmann2001advances} was applied to improve the conditioning of the reconstruction. The multi-coil data was compressed to a single channel prior to all reconstructions, see Supporting Information Section 1 for more details. The regularization parameter $\lambda_R$ was chosen according to $\lambda_R \sim 1/M$ and the data $\bv{s}_t$ were scaled by the norm of the density-compensated k-space data in order to obtain consistent values between experiments. The reconstruction parameters were determined with a heuristic parameter search (see Supporting Information Section 3). We refer to \autoref{tabel:experimentdetails} for all parameter settings and to Supporting Information Section 4 and the Supporting Information in \citep{huttinga2020mr} for more implementation details.

\subsection{Experiment 1: 2D+t {\it\bfseries in-vivo} respiratory motion reconstructions from abdominothoracic data}
\label{section:2dtreconstructionsmethods}
In the first experiment, a reference image and motion-fields were reconstructed from the same 2D+t data acquired during 20 seconds of free-breathing. The reference image was reconstructed from the end-inhale bin after phase-binning based on the self-navigation signal of $\bv{k}=0$ values per readout (denoted as $k_0$-values), see Supporting Information Section 2 for more details. The motion-fields were reconstructed at 40.8 Hz, i.e. 24.5 ms/frame, by assigning every 5 consecutive non-overlapping spokes to one dynamic. The low-rank model \eqref{eq:lowrankmotionmodel} was employed with $R=3$, yielding motion-fields with rank $\le 3$. Additional relevant reconstruction and acquisition parameters can be found in \autoref{tabel:experimentdetails}.

The motion-fields were analyzed by comparison with a time-resolved compressed sensing 2D+t reconstruction (CS2Dt) on the same free-breathing data, and by means of the Jacobian determinant. For the comparison with CS2Dt, the MR-MOTUS reference image was warped with the reconstructed motion-fields to obtain a dynamic image sequence as follows. First, the motion-fields
are interpolated to the same spatial resolution as the image reconstruction using cubic interpolation. Second, the forward model \eqref{eq:signalmodel} was evaluated on a Cartesian k-space grid using the reconstructed motion-fields $\bv{D}_t$. Finally, an inverse Fourier transform was performed to obtain one image per dynamic. The CS2Dt was reconstructed at a temporal resolution of 122.5 ms/frame by assigning every 25 consecutive non-overlapping spokes to one dynamic, and was performed with the BART toolbox \citep{uecker2015berkeley} using spatial $L_1$-wavelet and temporal total variation regularization. The temporal resolution of the CS2Dt was chosen as an integer multiple of the MR-MOTUS resolution to allow comparison at the coarser CS2Dt temporal resolution. The comparison was performed by means of the relative error norm (REN). The REN between vectors $\bv{a},\bv{b}$ was defined as $\text{REN}(\bv{a},\bv{b})=\frac{\lVert \bv{a} - \bv{b} \rVert}{\lVert \bv{b} \rVert}$.

\subsection{Experiment 2: 3D+t {\it\bfseries in-vivo} respiratory motion reconstructions from abdominothoracic data}
\label{section:methodsresp3d}
In the second experiment we considered the targeted case for MR-guided radiotherapy: a reference image and motion-fields were reconstructed from 3D+t data acquired during 33 seconds of free-breathing. The targeted high temporal resolution does not allow for a straightforward validation by comparison with dynamic 3D image reconstruction. For validation purposes, we therefore compared MR-MOTUS with respiratory-resolved image reconstruction by performing both reconstructions on respiratory-sorted data. 

Finally, we performed 3D+t time-resolved motion reconstruction to demonstrate the ability to reconstruct motion at high temporal resolution from time-resolved k-space data. The reference image for both reconstructions was reconstructed from the end-inhale bin after phase-binning based on the $k_0$-value per readout (see Supporting Information Section 2), and the low-rank model \eqref{eq:lowrankmotionmodel} was employed with $R=3$. See \autoref{tabel:experimentdetails} for all reconstruction and acquisition parameters.

For the respiratory-resolved reconstructions phase-binning was performed in 20 equal-sized bins based on the $k_0$-value per readout. The images were independently reconstructed for each bin using 28 iterations of CG-SENSE \citep{pruessmann2001advances}. The motion-fields were reconstructed over all bins simultaneously with low-rank MR-MOTUS by solving \autoref{eq:inverseproblemspatiotemporal_compressed} with 20 dynamics. The quality of the MR-MOTUS reconstruction was assessed by means of the Jacobian determinant and by comparison with the respiratory-resolved image reconstruction. For the latter, a reference image was warped with the reconstructed motion-fields to obtain a dynamic image sequence, as described in \cref{section:2dtreconstructionsmethods}, and the two image sequences were compared in terms of REN. The reference image that was warped using the MR-MOTUS motion-fields was selected as the end-inhale phase of the respiratory-resolved image reconstruction (motion state \#10) in order to reduce effects of image intensity, image quality, or contrast differences on the comparison of the two image sequences.

For the time-resolved 3D+t reconstructions, motion-fields were reconstructed at 7.6 Hz, i.e. 132 ms/frame, by assigning every 30 consecutive non-overlapping spokes to one dynamic. The reconstructions were analyzed by means of the Jacobian determinant and the average motion of the kidney was compared between the time-resolved and respiratory-resolved MR-MOTUS reconstructions. This motion was computed as the mean of the displacements over a manually segmented mask of the right kidney. For comparison between respiratory-resolved and time-resolved, the motion magnitudes of each respiratory bin in the respiratory-resolved reconstruction were assigned to the original, time-resolved, spoke indices that were sorted into that particular bin.

\subsection{Experiment 3: 3D+t {\it\bfseries in-vivo} head-and-neck motion reconstructions}
\label{section:methodshead3d}

With the third experiment, 3D+t motion-fields were reconstructed from data acquired during head-and-neck motion. The subject was instructed to hold still in position 1 during the first 70 seconds of the acquisition, then move to position 2 and hold still for 70 seconds, then move freely for 40 seconds, and finally hold still afterwards in position 3 for 70 seconds. Data acquired in position 1 was used to reconstruct a reference image, data acquired during movement from position 2 to position 3 was used to reconstruct motion-fields, and position 2 and 3 were used as fully-sampled "checkpoints" to serve as validation; the beginning and end of the dynamic motion reconstruction should respectively coincide with positions 2 and 3. To verify this, the reference was warped with the reconstructed motion-fields as described in \cref{section:2dtreconstructionsmethods}, and the first and last dynamic of the resulting image sequence were visually compared with the fully-sampled checkpoints. As a second analysis, the mean and standard deviation of the determinant of the Jacobian were computed for all dynamics, over all voxels within the body. The latter were determined by a threshold on the magnitude of the signal per voxel. 
The low-rank motion model was employed with $R=6$ to accommodate the head-and-neck motion which includes rotations in multiple planes. The motion-fields were reconstructed at a temporal resolution of 9.3 Hz, i.e. 108 ms/frame, by assigning every 20 consecutive non-overlapping spokes to one dynamic. Additional reconstruction and acquisition parameters can be found in \autoref{tabel:experimentdetails}.

\section{Results}

\subsection{Experiment 1: 2D {\it\bfseries in-vivo} respiratory motion reconstructions from abdominothoracic data}
\label{section:2dtreconstructionsresults}
\begin{figure}[h!]
    \centering
    \includegraphics[width=\textwidth]{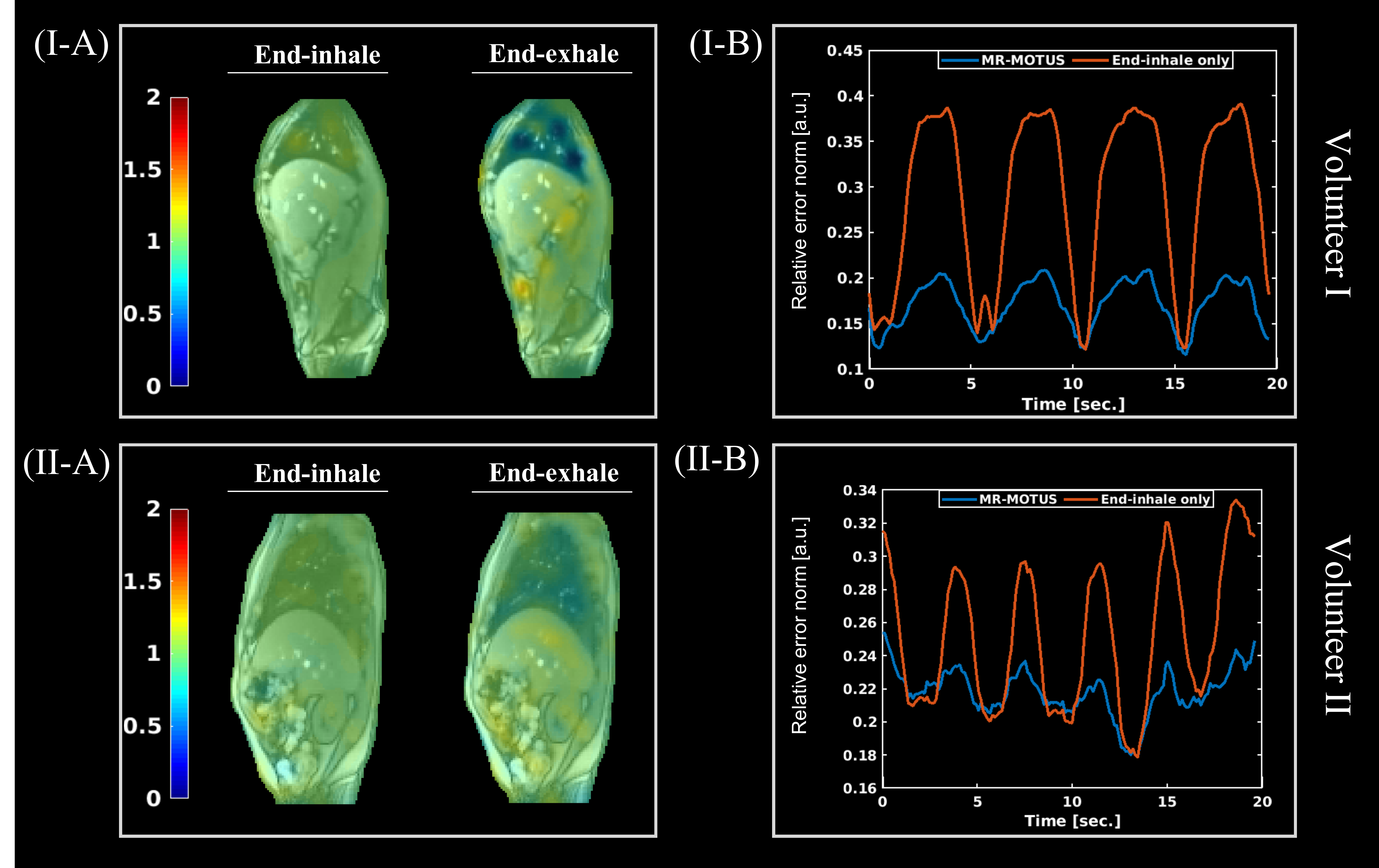}
    \caption{A) Jacobian determinants of the reconstructed motion-fields in end-inhale (left) and end-exhale (right). The first end-exhale and second end-inhale positions were selected from all dynamics for this visualization. B) Relative error norm (REN) between MR-MOTUS warped reference images and CS2Dt reconstruction over all dynamics (blue), and a baseline REN between the fixed MR-MOTUS end-exhale warped reference image and CS2Dt. The top row (I) shows the results for volunteer 1, whereas the bottom row (II) shows the results for volunteer 2. The comparison is also visualized in \ref{video:TR2DtvsCS2Dt}, and the reconstructed motion-fields decomposed in the low-rank model components are visualized in \ref{video:TR2DtMIO} and \ref{video:TR2DtMIO_VOL2}.}
    \label{fig:TR2DtvsCS2DtDeterminantNRMSE}
\end{figure}

The time-resolved 2D respiratory motion was reconstructed with 40.8 motion-fields-per-second. The Jacobian determinant and the comparison with CS2Dt is shown in \ref{fig:TR2DtvsCS2DtDeterminantNRMSE}. The visual comparison with 2D+t compressed sensing image reconstruction corresponding to \ref{fig:TR2DtvsCS2DtDeterminantNRMSE}B is shown in \ref{video:TR2DtvsCS2Dt}. It can be observed that good agreement is obtained for most phases of the respiratory cycle, with a small mismatch in end-exhale in the upper back near the spine-liver interface. The Jacobian determinants show small deviations from unity within the organs (green), and compression in the lungs (blue) except for the arteries. The qualitative results are supported by the quantitative results in \ref{fig:TR2DtvsCS2DtDeterminantNRMSE}B, which show that the warped MR-MOTUS images considerably reduce the REN.   

The warped reference images corresponding to the reconstructed motion-field, overlayed with the motion-field are shown in \ref{video:TR2DtMIO} and \ref{video:TR2DtMIO_VOL2}. Moreover, these show the decomposition in the reconstructed low-rank components. For volunteer 1, the first two components show pseudo-periodic temporal behaviours, and the first is most prominent in magnitude. Both components show realistic movement of organs such as the liver and kidney, but also small unrealistic motion in the spine near the liver in end-exhale. Interestingly, the third component shows a temporal behavior with a slight drift upwards, and the corresponding spatial motion-field indicates a global rotation. Similar movement can also be observed in the ground-truth CS2Dt reconstruction in \ref{video:TR2DtvsCS2Dt}. This movement could be caused by relaxation of the gluteus maximus muscle in the upper leg and buttocks. Similar motion patterns can be observed in \ref{video:TR2DtMIO_VOL2} for volunteer 2, but the global rotation is less pronounced in the ground-truth CS2Dt reconstruction.

\subsection{Experiment 2: 3D {\it\bfseries in-vivo}
respiratory motion reconstructions from abdomen/thorax data}
\label{section:resultsresp3d}
\begin{figure}[h!]
    \centering
    \includegraphics[width=\textwidth]{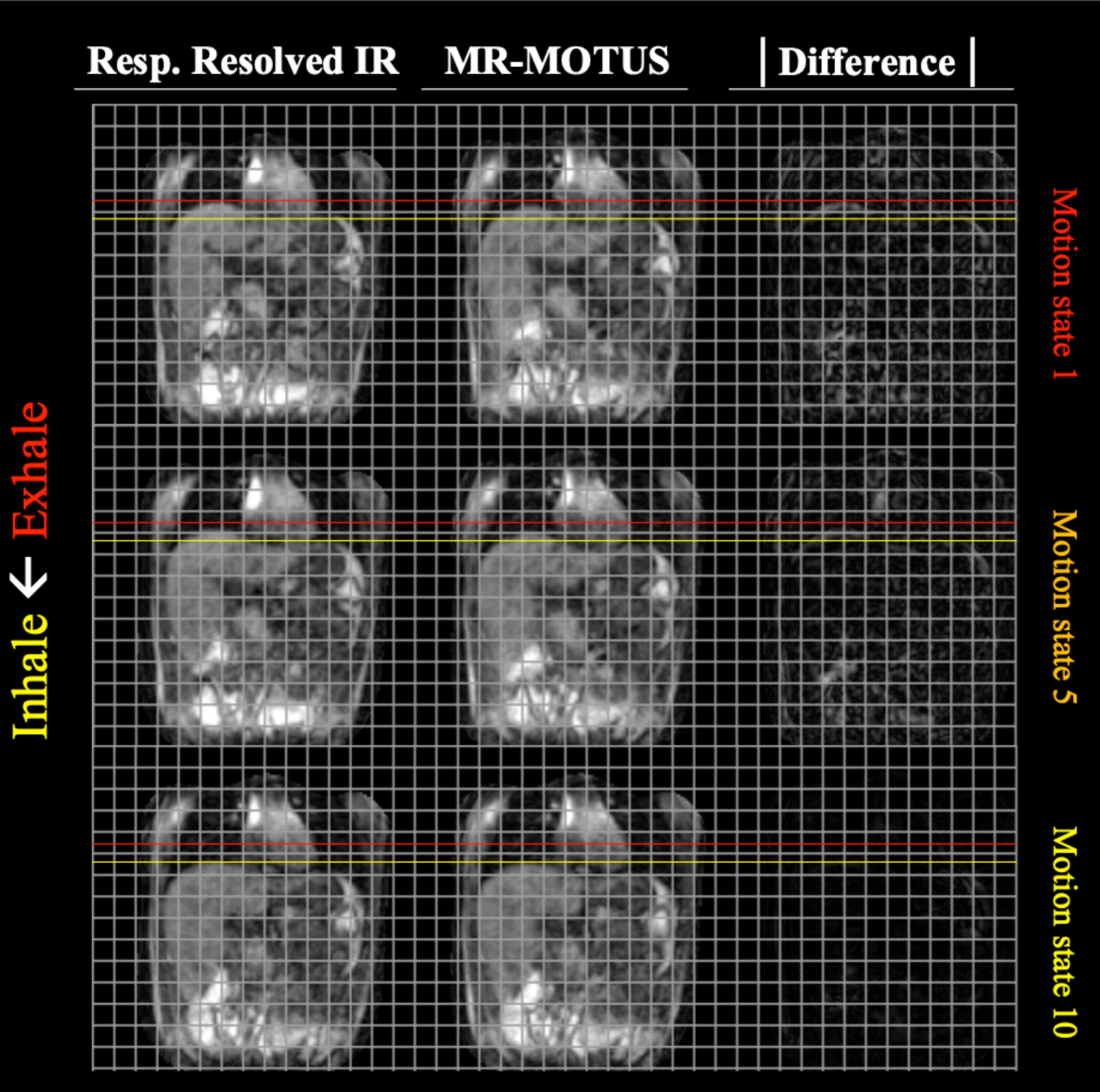}
    \caption{Respiratory-resolved image reconstruction (Resp. Resolved IR, left), MR-MOTUS warped reference image (middle), and pixel-wise absolute difference between the two reconstructions (right), as mentioned in \cref{section:methodsresp3d} and \cref{section:resultsresp3d}. The red and yellow horizontal lines indicated respectively end-exhale and end-inhale positions. A video corresponding to this figure of volunteer 1 is provided in \ref{video:RRMRMOTUS_vs_RRIR}. A similar video for volunteer 2 is provided in \ref{video:RRMRMOTUS_vs_RRIR_VOL2}.}
    \label{fig:RRMRMOTUS_vs_RRIR}
\end{figure}

\begin{figure}[h!]
    \centering
    \includegraphics[width=\textwidth]{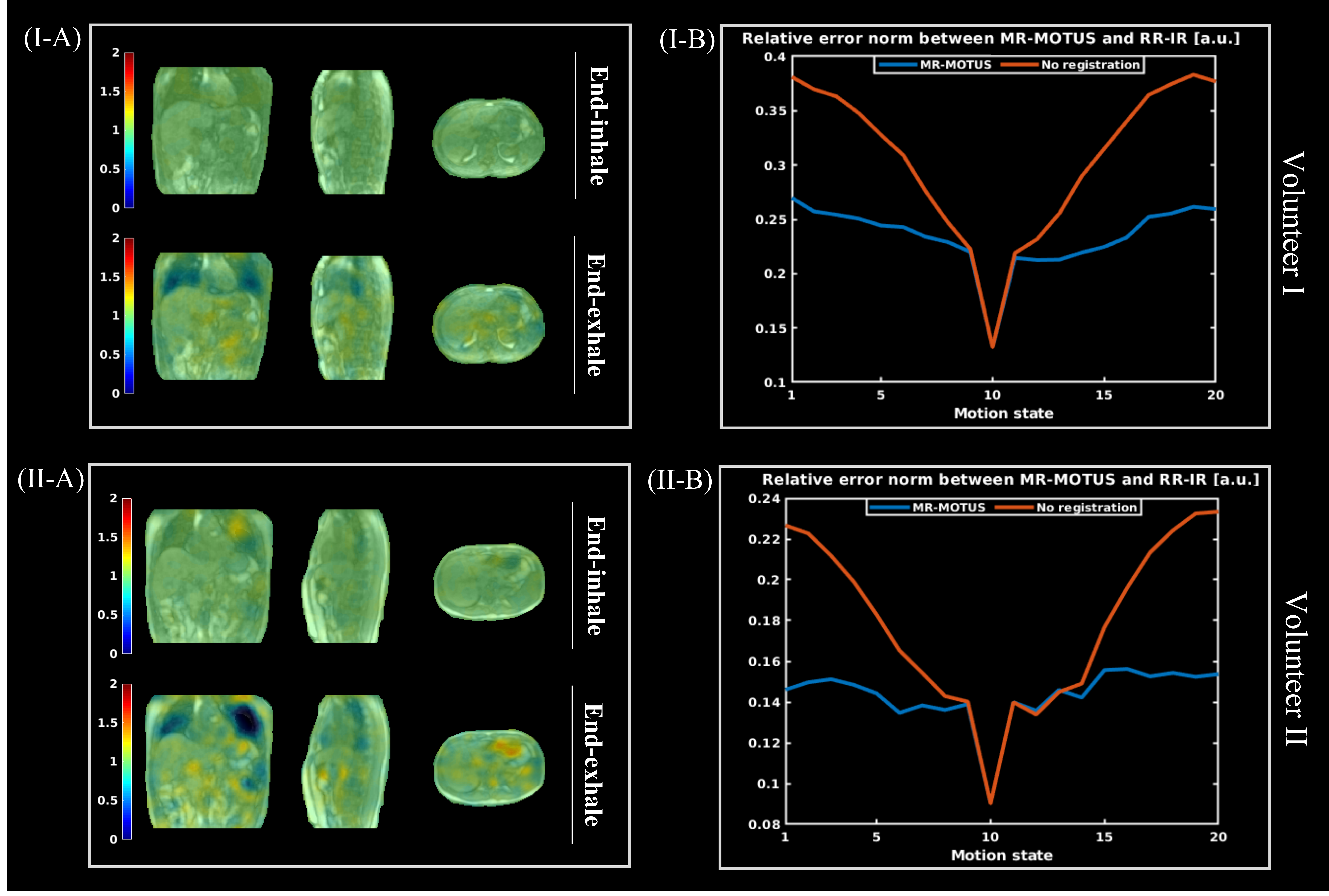}
    \caption{A) Jacobian determinants of the reconstructed respiratory-resolved motion-fields in end-inhale (top) and end-exhale (bottom). The first end-exhale and second end-inhale positions were selected from all dynamics for this visualization. B) Relative error norm (REN) with respiratory-resolved image reconstruction (RR-IR) for every motion state. The blue graph indicates the REN between MR-MOTUS and respiratory-resolved image reconstruction. The orange graph indicates a baseline comparison between the (fixed) end-inhale image of the MR-MOTUS reconstruction and the (dynamic) respiratory-resolved image reconstruction. The sharp peak is caused by taking the 10th dynamic as the reference image for this comparison. The top row shows the results for volunteer 1, and the bottom row shows the results for volunteer 2. Videos corresponding to the comparisons in (B) are provided in \ref{video:RRMRMOTUS_vs_RRIR} and \ref{video:RRMRMOTUS_vs_RRIR_VOL2}.}
    \label{fig:RRMRMOTUS_DeterminantNRMSE}
\end{figure}

The comparison between MR-MOTUS and respiratory-resolved image reconstruction is shown in
\ref{fig:RRMRMOTUS_vs_RRIR}, \ref{fig:RRMRMOTUS_DeterminantNRMSE},  \ref{video:RRMRMOTUS_vs_RRIR} and \ref{video:RRMRMOTUS_vs_RRIR_VOL2}. It can be observed that good visual agreement is obtained between the two reconstructions for both volunteers. This is especially visible from the position of the top of the liver dome. The Jacobian determinants of the reconstructed motion-fields are shown in \ref{fig:RRMRMOTUS_DeterminantNRMSE}A. The lungs show compression (blue), except for the arteries, and small deviations from unity can be observed in the rest of the body. Deviation from unity can be observed at the spine-liver interface, where a large volumetric compression is reconstructed. We expect this is related to the attachment of liver tissue to the spine during exhalation. The quantitative comparison in \ref{fig:RRMRMOTUS_DeterminantNRMSE}B shows best agreement at motion state 10 (inhale) and worst agreement in motion state 19 (exhale). The sharp peak at motion state 10 can be explained by the fact that we took motion state 10 as the reference image to compute the warped reference images for MR-MOTUS. The warped reference images reconstructed from the respiratory-sorted data, overlayed with the motion-field, are visualized for both volunteers in \ref{video:RR_RankDecomp} and \ref{video:RR_RankDecomp_VOL2}. Moreover, these show the decomposition in the reconstructed low-rank components. For both volunteers the first component shows a pseudo-periodic behavior in time and is most prominent in magnitude; the other components make only minor contributions. These large contributions of pseudo-periodic components could be due to the periodicity assumption underlying the respiratory-sorting. Small unrealistic motion can be observed for volunteer 1 at the spine-liver interface and at the back of the spine, similar to the 2D reconstructions. Additionally, a small rotating motion can be observed in the motion-field for volunteer 1 at the interface with the rib cage in the coronal slice on the bottom right. We expect the latter is caused by a combination of the volume-preservering regularization and the inability of the motion model to resolve the sliding motion that is present in this area. 

\begin{figure}[h!]
    \centering
    \includegraphics[width=\textwidth]{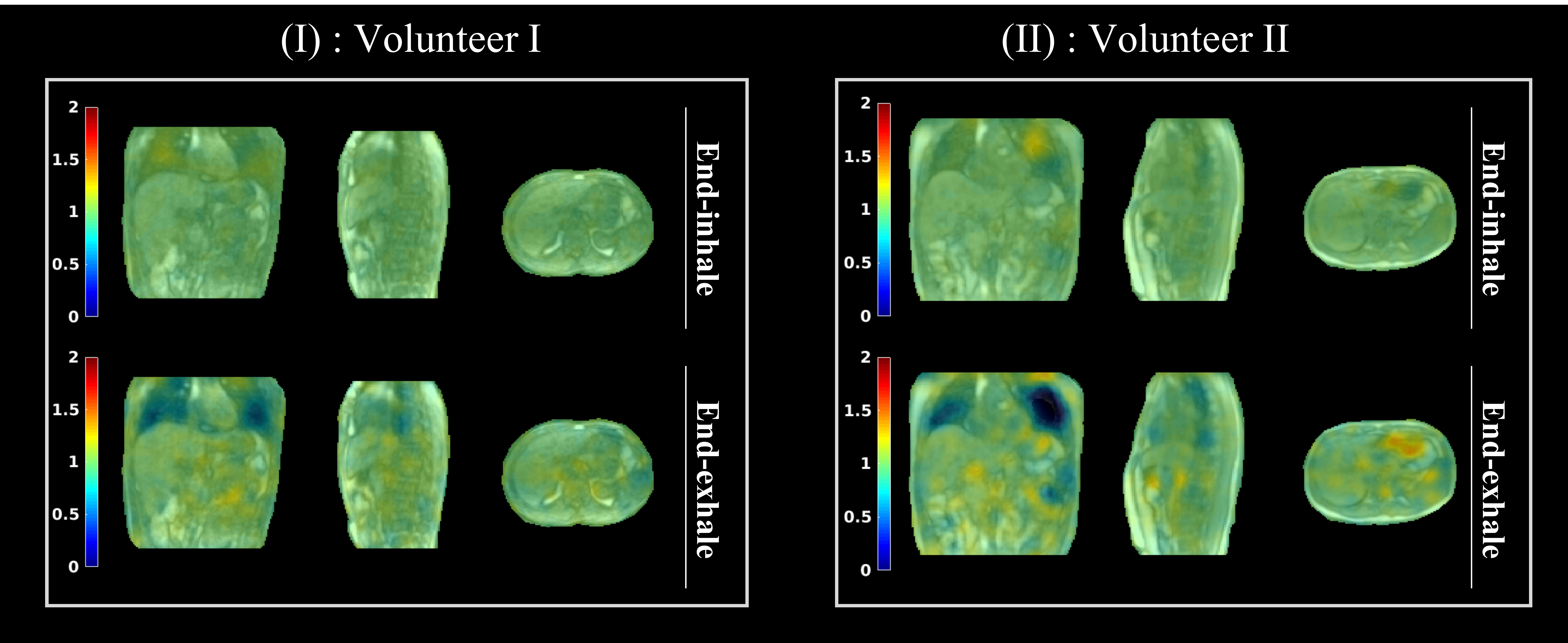}
    \caption{Jacobian determinants of the reconstructed time-resolved motion-fields in end-inhale (top) and end-exhale (bottom). The left figure shows the results for volunteer 1 and the right figure the results for volunteer 2. Videos corresponding to the reconstructions in this figure are provided in \ref{video:TR3Dt_Resp_MRMOTUS} and \ref{video:TR3Dt_Resp_MRMOTUS_VOL2}. \\\\}
    \label{fig:TR_DeterminantNRMSE}
\end{figure}

\begin{figure}[h!]
    \centering
    \includegraphics[width=\textwidth]{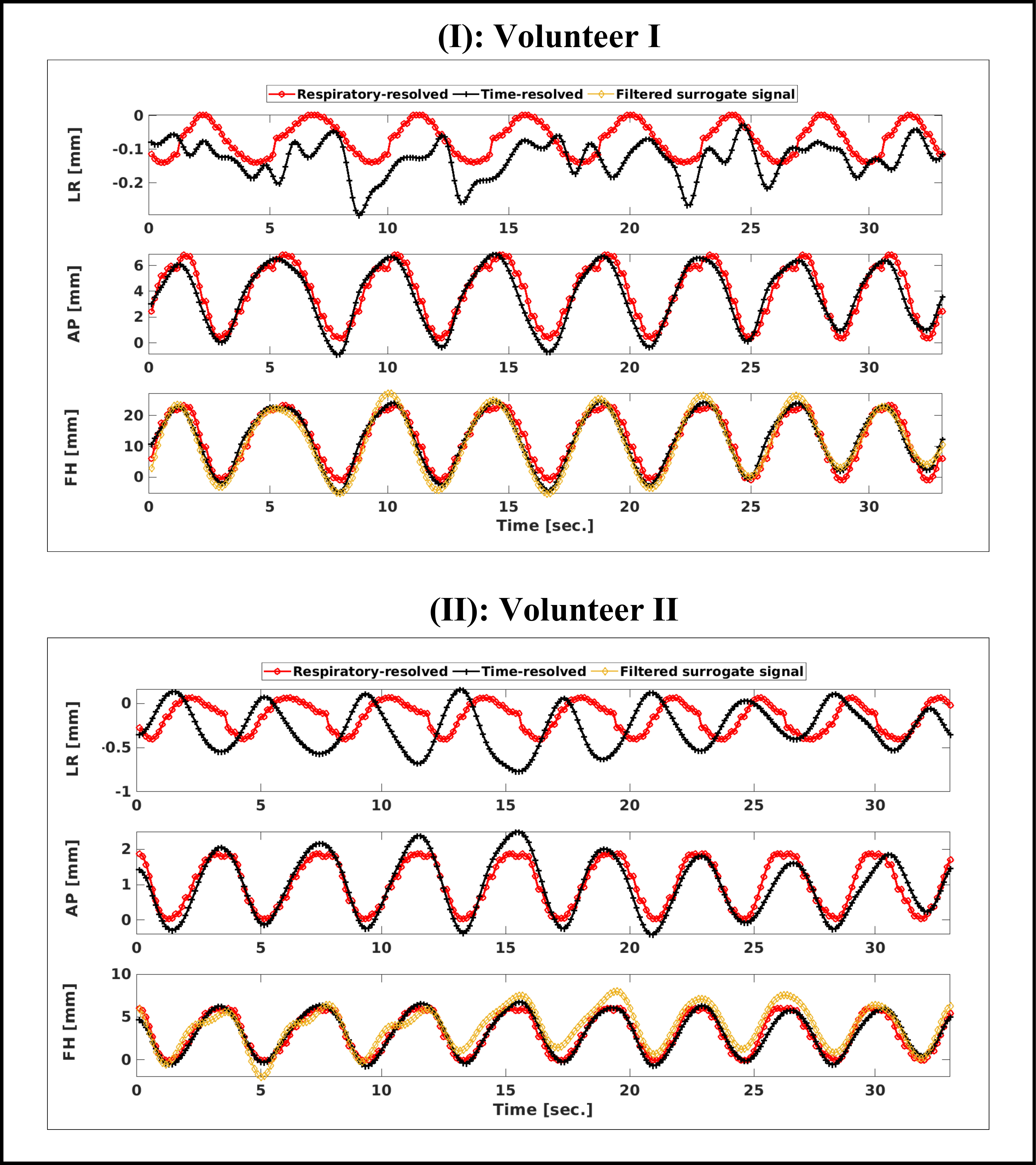}
    \caption{This figure shows the average motion of the right kidney over time, for both the respiratory-resolved and the time-resolved MR-MOTUS reconstructions mentioned in \cref{section:methodsresp3d} and \cref{section:resultsresp3d}. The respiratory-resolved MR-MOTUS reconstruction was projected back on the time axis, as described in \cref{section:methodsresp3d} and \cref{section:resultsresp3d}. The average motion magnitudes were computed over a manually segmented mask of the right kidney. Videos of reconstructions corresponding to these figures are provided in \ref{video:TR3Dt_Resp_MRMOTUS} and \ref{video:TR3Dt_Resp_MRMOTUS_VOL2}.}
    \label{fig:3DTR_KidneyMotion}
\end{figure}

The time-resolved 3D respiratory motion was reconstructed with 7.6 motion-fields-per-second. The warped reference images reconstructed from the time-resolved data, overlayed with the motion-field, are visualized for both volunteers in \ref{video:TR3Dt_Resp_MRMOTUS} and \ref{video:TR3Dt_Resp_MRMOTUS_VOL2}. Similar motion is obtained as with the respiratory-sorted data, but the reconstructed motion components are now similar in magnitude. All components show pseudo-periodic temporal behavior, and the first component of volunteer 1 indicates a small drift. Similar to the respiratory-resolved reconstructions, small unrealistic motion at the spine-liver interface and anterior side of the spine can be observed for volunteer 1. Additionally, the same small rotation can be observed near the rib cage in the bottom right of the coronal slice. The Jacobian determinants of the reconstructed motion-fields are shown in \ref{fig:TR_DeterminantNRMSE}. Similar patterns can be observed in end-exhale as for the respiratory-resolved motion reconstructions. Interestingly, the end-inhale image for volunteer 1 shows a small expansion in the lungs, possibly indicating that a deeper inhale than the reference image was reconstructed while the reference image was obtained using respiratory-sorting on end-inhale. Finally, the comparison between the average kidney motion in the time-resolved and respiratory-resolved MR-MOTUS reconstructions is visualized in \ref{fig:3DTR_KidneyMotion}. The phase of the reconstructions are most similar in feet-head (FH) and anterior-posterior (AP), while in left-right (LR) different patterns can be observed. However, it should be noted that the motion in FH and AP is two orders of magnitude higher than in LR. The motion magnitude is similar for both reconstructions, but the respiratory-resolved reconstruction shows a constant amplitude over time since it only reconstructs an average breathing cycle. The time-resolved reconstruction shows changing motion amplitudes over time. The phase difference between the two reconstructions may be explained by imperfect respiratory-sorting.

\subsection{Experiment 3: 3D {\it\bfseries in-vivo} head-and-neck motion reconstructions}
\label{section:resultshn3d}
\begin{figure}[h!]
    \centering
    \includegraphics[width=\textwidth]{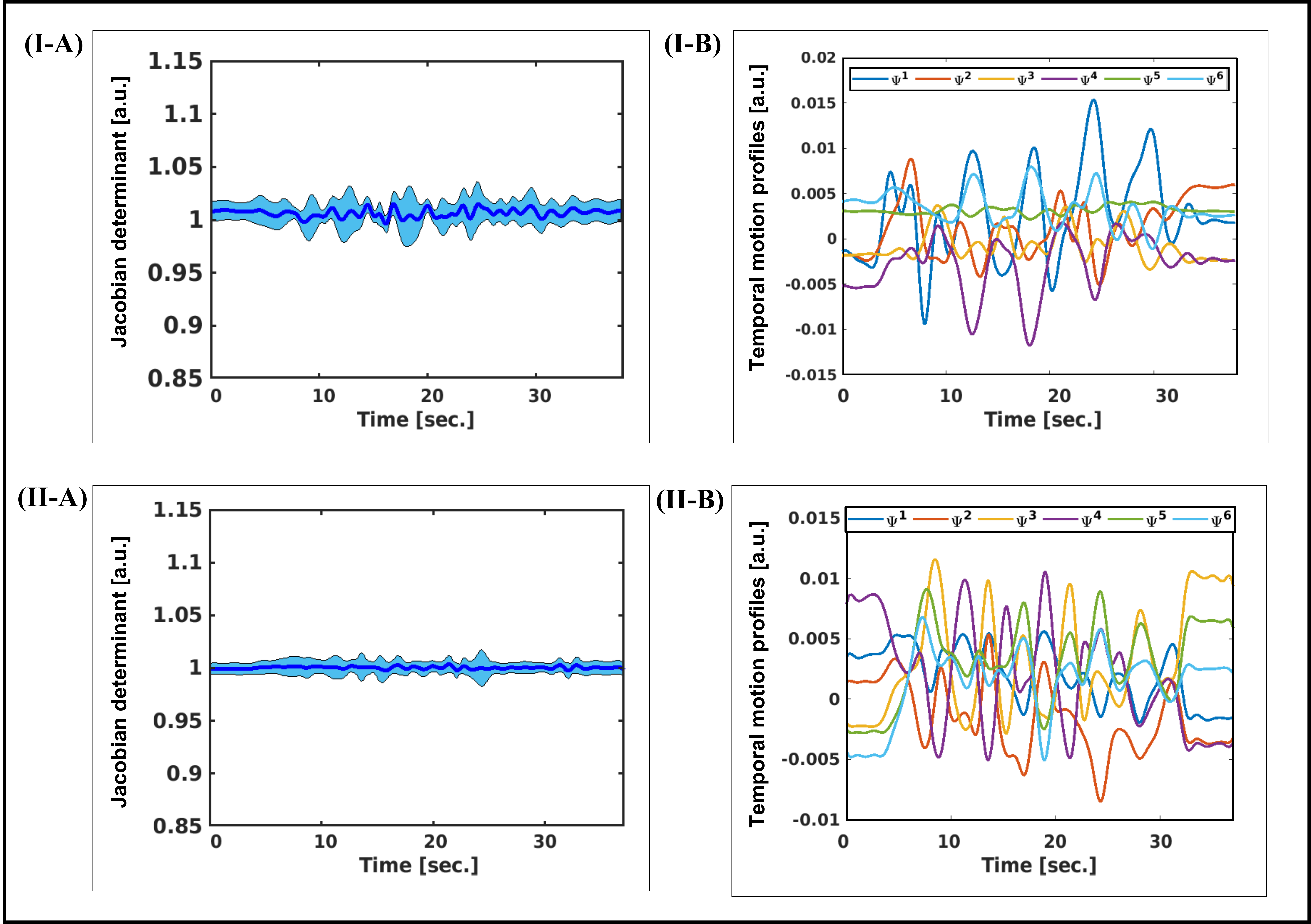}
    \caption{This figure corresponds to the head-and-neck reconstructions in \cref{section:methodshead3d} and \cref{section:resultshn3d}. A) The mean (solid line) and standard deviation (shaded area) of the Jacobian determinants of the reconstructed motion-fields over time. B) The reconstructed temporal profiles $\bm{\Psi}^i$, scaled by the norm of the corresponding $\bm{\Phi}^i$ to be able to compare their magnitudes. The top row and bottom rows respectively show the results for volunteer 1 and 2. Videos corresponding to the reconstructions in these figures are provided in \ref{video:TR3Dt_HN_MRMOTUS} and \ref{video:TR3Dt_HN_MRMOTUS_VOL2}.}
    \label{fig:3D_HN_ImageDetOverlay}
\end{figure}

\begin{figure}[h!]
    \centering
    \includegraphics[width=0.55\textwidth]{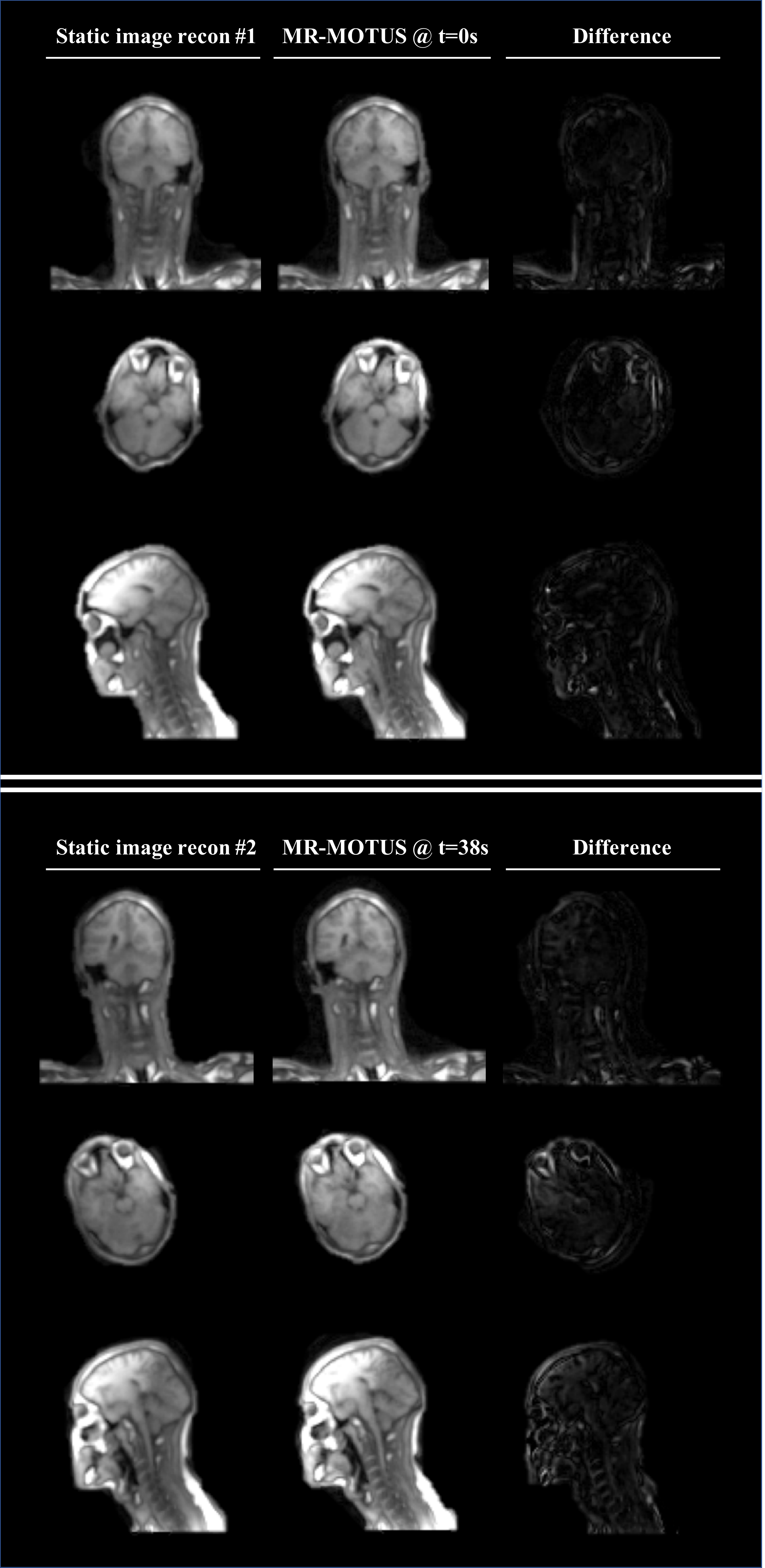}
    \caption{This figure shows the checkpoint validation for the head-and-neck reconstructions of volunteer 2, as mentioned in \cref{section:methodshead3d} and \cref{section:resultshn3d}. The left columns shows the fully-sampled checkpoint image, the middle column shows the MR-MOTUS warped reference images and the right column shows the absolute pixel-wise difference. The top part corresponds to the comparison with the checkpoint acquired right before the start of the motion, and the bottom part corresponds to the checkpoint acquired right after the start of the motion. A video corresponding to this figure is provided in \ref{video:TR3Dt_HN_MRMOTUS_VOL2}. A similar video for volunteer 1 is provided in \ref{video:TR3Dt_HN_MRMOTUS}.}
    \label{fig:3D_HN_checkpoint}
\end{figure}

The time-resolved 3D head-and-neck motion was reconstructed with 9.3 motion-fields-per-second.  The MR-MOTUS warped reference images from 3D data acquired during head-and-neck motion are visualized for both volunteers in \ref{video:TR3Dt_HN_MRMOTUS} and \ref{video:TR3Dt_HN_MRMOTUS_VOL2}. Clearly, rigid motion-fields are reconstructed within the skull, and non-rigid motion-fields at the neck. \ref{fig:3D_HN_ImageDetOverlay} shows the Jacobian determinants of the reconstructed motion-field over time (A), and the reconstructed temporal components (B) for both volunteers. The Jacobian determinant is close to 1 over the whole reconstructed time, with slightly more deviations for volunteer 1. These can be attributed to larger and more irregular motion than volunteer 2. The temporal components are relatively flat at the start and the end, corresponding to the static begin and end positions. The more extreme motion of volunteer 1 can also be observed from the larger magnitudes of the temporal components and from \ref{video:TR3Dt_HN_MRMOTUS}. \ref{fig:3D_HN_checkpoint} shows the checkpoint validation for volunteer 2. It can be observed that good agreement is obtained between the fully-sampled checkpoint images and the MR-MOTUS reconstructions.
 
\section{Discussion}
We have previously introduced MR-MOTUS \citep{huttinga2020mr}, a framework to estimate motion directly from minimal k-space data and a reference image by exploiting spatial correlation in internal body motion. In this work, we introduce low-rank MR-MOTUS: an extension of MR-MOTUS from 3D to 3D+t reconstructions in a realistic experimental setting, where both reference image and motion-fields are reconstructed from data acquired during free-breathing. Low-rank MR-MOTUS employs a low-rank motion model that constrains the degrees of freedom in space and time, thereby reducing memory consumption and functioning as a regularization in both space and time. It was demonstrated that the proposed method can reconstruct high quality whole-body 3D motion-fields with a temporal resolution of more than 7.6 motion-fields-per-second, while showing consistency with static, respiratory-resolved and time-resolved image reconstructions. Prospectively undersampled data were acquired with a non-Cartesian trajectory and multi-channel receivers, thereby bridging the gap towards clinical application.

The ability of the proposed framework to estimate time-resolved rather than respiratory-resolved motion is promising as it allows to characterize drifts and intra and inter-cycle breathing patterns. This is in contrast with respiratory-resolved methods that require sorting to obtain suitable images \citep{breuer2018stable,deng2016four,han2017respiratory,cai2011four,feng2016xd,stemkens2016image,Feng2020,rank20174d,jiang2018motion}. The sorting effectively results in (a motion model for) average breathing motion, which may have trouble capturing drifts and inter-cycle variations. Some works have been proposed to reconstruct time-resolved MR-images without the need of retrospective sorting. However, the reported temporal resolution was too low \citep{yuan2019fast,ong2019extreme, king2012thoracic}, or the FOV was too small \citep{fu2017high,burdumy2017one}. The time-resolved motion estimation of low-rank MR-MOTUS in combination with an MR-LINAC can be particularly beneficial for MR-guided radiotherapy; the (retrospective) reconstruction of 3D+t time-resolved tumor and organs-at-risk motion during treatment can be used for accurate dose accumulation \citep{cai20153d}, allowing for an accurate assessment of the treatments. 

The resulting motion model explicitly separates a high-dimensional static spatial component from a low-dimensional dynamic temporal component. The low-dimensional compression of the dynamic behavior could be exploited to reduce the number of parameters and reconstruction times of future real-time reconstructions, analogously to recently proposed approaches in \citep{sbrizzi2019acquisition,mcclelland2017generalized,stemkens2016image,huttinga2020realtime}. Our method could thereby form the basis for future work on real-time MR-based motion estimation, where reconstructions are performed on-the-fly to track tumor and organs-at-risk motion. 

Low-rank models in the context of motion estimation have been investigated before in several works, most of which retrospectively perform compression to a low-rank model using principal component analysis \citep{zhang2007patient,stemkens2016image,mishra2014initial,cai20153d}. Others decouple the motion-fields into spatial components and temporal components based on surrogate signals \citep{Low2005,mcclelland2013respiratory,mcclelland2017generalized}. The approach in this work is different in the sense that it explicitly and a-priori enforces a structure that yields low-rank motion-fields, and does not assume dependence on surrogate signals for the motion model. Similar approaches have been studied in the context of image reconstruction \citep{ong2019extreme,fu2017high,lingala2011accelerated,Liang2007,Zhao2012,Zhao2010,Haldar2010}. 

This work includes some limitations and assumptions that should be addressed. Both the respiratory-resolved and time-resolved 3D respiratory motion reconstructions in \cref{section:methodsresp3d} and \cref{section:resultsresp3d} look realistic in general. Yet, small unrealistic motion is reconstructed near discontinuities in the true motion-fields that are present near sliding or attaching/detaching organ surfaces. This can be observed in for example \ref{video:TR2DtMIO}, at the spine/liver interface in end-exhale; the increase in compression due to the attaching tissue is compensated by unrealistic movement in the spine. This could possibly be resolved with region-specific \citep{Delmon2013} or non-parametric motion models \cite{Fu2018}, but is beyond the scope of this work.

Another point of improvement is the compression of multi-channel data to a single channel (Supporting Information Section 1). Contrary to standard coil compression techniques, the aim of the compression in this work is homogeneous coil sensitivity. Consequently, this compression is suboptimal in terms of SNR \citep{Roemer1990,Kellman2005}. \ref{fig:2D_snr_comparison} analyzes the loss between between a Roemer coil combination \citep{Roemer1990} and the proposed coil compression on the 2D data. This shows an SNR loss factor between 1.5 and 2.5 in most of the body, which increases towards the boundary of the body. Good results were obtained with the coil compression introduced in this work, but more advanced techniques could possibly be used to improve the SNR after the compression.

The last point of improvement is the validation of time-resolved 3D motion-fields. In general this is not straightforward, and we considered three viable options for this: (1) {\it in-silico} with a digital phantom, (2) with an MR-compatible motion phantom, and (3) {\it in-vivo} with respiratory-resolved image reconstruction. We have opted for the third option, since this was considered the closest to a practical use-case. The {\it in-silico} validation does not consider real acquisition-related data corruption (e.g. eddy currents, flow effects), and can, in case of e.g. the XCAT phantom \citep{Segars2010}, yield unstable motion-fields \citep{Eiben2020}. MR-compatible motion phantoms, although useful for proof-of-principle validations, have limitations regarding the representation of realistic {\it in-vivo} anatomies.

The intended application of MR-MOTUS is MR-guided radiotherapy, possibly in real-time. However, the current reconstruction times in MATLAB on a desktop workstation are around 4 minutes for 2D+t with 40 motion-fields/second, around 6 minutes for the respiratory-resolved 3D reconstruction, and around 50 minutes for 3D+t time-resolved respiratory motion with 7.6 motion-fields/second. Hence, the current implementation of the method is not directly applicable for real-time processing, but reconstruction times may be reduced with a different programming language, improved hardware, GPU-accelerations or deep learning.

\section*{Conclusion}
We have introduced low-rank MR-MOTUS, an extension of MR-MOTUS, that allows to retrospectively reconstruct whole-body time-resolved 3D+t motion-fields from prospectively undersampled k-space data and one reference image. Reconstructions were performed for 2D/3D respiratory motion and 3D head-and-neck motion. A temporal resolution of more than 7.8 motion-fields-per-second was obtained, and the motion-fields were consistent with image reconstructions. For MR-guided radiotherapy, the time-resolved 3D motion-fields could be used to reconstruct the respiratory-motion-compensated accumulated dose during the treatment. Furthermore, the explicit decomposition of motion-fields in static and dynamic components could form the basis for future work towards real-time MR-guided radiotherapy.


\clearpage

\section*{Figures and Tables}

\begin{table*}
    \centering
    \resizebox{\textwidth}{!}{\begin{tabular}{lllll}
        \multicolumn{4}{c}{\bfseries\Large Acquisition details} \\ \addlinespace \hline \hline {\bf Parameter} & \multicolumn{1}{l}{\bf 2D respiratory motion } & \multicolumn{1}{l}{\bf 3D respiratory motion } & \multicolumn{1}{l}{\bf 3D head-and-neck motion } \\  \hline 
         FOV [m] & $0.50\times0.50 \times 0.01$ & $0.44\times0.44 \times 0.44$ & $0.38 \times 0.38 \times 0.38$\\ 
         
         \rowcolor{lightgray} Acquisition matrix size  &  $164\times164\times1$ & $146\times146\times146$ & $126\times126\times126$ \\ 
         
         Spatial acq. resolution [mm] & $3.00\times 3.00\times 10.00 $ & $3.00\times3.00\times3.00$ & $3.00\times3.00\times3.00$ \\ \rowcolor{lightgray}

        Repetition time [ms] & 4.90  & 4.40 & 5.40  \\ 
        
        Echo time [ms] & 2.30  &1.80 & 2.30 \\ \rowcolor{lightgray}
        
        Flip angle [\textdegree] & 20 & 20 & 20\\
        
        Bandwidth [Hz]  & 298.72 & 541.48 & 284.73 \\ \rowcolor{lightgray}
        
        Trajectory & 2D golden-angle radial & 3D golden-mean radial kooshball & 3D golden-mean radial kooshball \\

        Pulse sequence & 2D SPGR & 3D SPGR & 3D SPGR \\ \rowcolor{lightgray}
        
        Coils (\#Channels) & Anterior + Posterior (24) & Anterior + Posterior (24) & Anterior + Posterior (24) \\ Scanner & Philips Ingenia 1.5T & Philips Ingenia 1.5T & Philips Ingenia 1.5T \\ \hline \\ \addlinespace \addlinespace
        \multicolumn{4}{c}{\bfseries\Large Reconstruction details} \\ \addlinespace \hline \hline {\bf Parameter} & \multicolumn{1}{l}{\bf 2D respiratory motion } & \multicolumn{1}{l}{\bf 3D respiratory motion } & \multicolumn{1}{l}{\bf 3D head-and-neck motion } \\  \hline 
        Motion model components & $R=3$ & $R=3$ & $R=6$ \\  \rowcolor{lightgray}
         Reference image resolution [mm] & $6.70\times 6.70\times 10.00 $ & $6.70\times 6.70\times 6.70 $ & $9.05\times 9.05\times 9.05 $ \\ Regularization parameter & $\lambda_R=1.5\cdot 10^1$ & $\lambda_R=1.5 \cdot 10^1$ & $\lambda_R=1.4 \cdot 10^3 $ \\ \rowcolor{lightgray}
         Number of iterations & 50 & 50 & 300 \\  Splines per spatial dimension  & 18 & 16 & 3 \\ \rowcolor{lightgray} 
         Splines in time &  1.28 / second & 8.25 / second & 5 / second \\ Temporal motion resolution & 40.8 Hz: 5 spokes / dynamic & 7.6 Hz: 30 spokes / dynamic & 9.3 Hz: 20 spokes / dynamic  \\  \rowcolor{lightgray}
         Reconstructed motion duration [s] & 20 & 33 & 40 \\ Reconstruction time & 4 minutes & 50 minutes & 2 hours  \\ \hline
    \end{tabular}}
    \caption{Details of the {\it in-vivo} experiments as described in \cref{section:2dtreconstructionsmethods}-\cref{section:methodshead3d}: the top half lists acquisition details, and the bottom half lists reconstruction details for the time-resolved experiments. For the respiratory-resolved reconstruction in \cref{section:methodsresp3d} the same parameters were used as listed in the `3D respiratory motion' column, but effectively resulted in a temporal motion resolution of about 5Hz, with 18062 spokes per dynamic, due to the sorting.}
\label{tabel:experimentdetails}
\end{table*}

\setcounter{figure}{0}
\clearpage
\noindent {\huge\bfseries Supplementary Figures}

\setsuppvideo
\begin{figure}[h!]
    \centering
    \caption{{\it This is an animated figure and should be viewed at \url{https://surfdrive.surf.nl/files/index.php/s/XcXY3cHETnmXtY0}.} 2D-t compressed sensing reconstruction (left), MR-MOTUS warped reference images (middle), and pixel-wise absolute differences between the two reconstructions (right), as mentioned in \cref{section:2dtreconstructionsmethods} and \cref{section:2dtreconstructionsresults}. The top row shows reconstructions for volunteer 1, and the bottom row for volunteer 2.}
    \label{video:TR2DtvsCS2Dt}
\end{figure}

\begin{figure}[h!]
    \centering        
    \caption{{\it This is an animated figure and should be viewed at \url{https://surfdrive.surf.nl/files/index.php/s/XcXY3cHETnmXtY0}.} MR-MOTUS warped reference images overlayed with reconstructed dynamic motion-fields from 2D time-resolved data, as mentioned in \cref{section:2dtreconstructionsmethods} and \cref{section:2dtreconstructionsresults}. The image shows a decomposition in the reconstructed components $\bm{\Phi}^i$ (spatial) and $\bm{\Psi}$ (temporal) for volunteer 1. For visualization purposes the components were scaled such that $\lVert \bm{\Phi}^i \rVert=1$.}
    \label{video:TR2DtMIO}
\end{figure}

\begin{figure}[h!]
    \centering        
    \caption{{\it This is an animated figure and should be viewed at \url{https://surfdrive.surf.nl/files/index.php/s/XcXY3cHETnmXtY0}.} MR-MOTUS warped reference images overlayed with reconstructed dynamic motion-fields from 2D time-resolved data, as mentioned in \cref{section:2dtreconstructionsmethods} and \cref{section:2dtreconstructionsresults}. The image shows a decomposition in the reconstructed components $\bm{\Phi}^i$ (spatial) and $\bm{\Psi}$ (temporal) for volunteer 2. For visualization purposes the components were scaled such that $\lVert \bm{\Phi}^i \rVert=1$.}
    \label{video:TR2DtMIO_VOL2}
\end{figure}

\begin{figure}[h!]
    \centering
    \caption{{\it This is an animated figure and should be viewed at \url{https://surfdrive.surf.nl/files/index.php/s/XcXY3cHETnmXtY0}.} Respiratory-resolved image reconstruction (Resp. resolved IR, left), MR-MOTUS warped reference images (middle), and pixel-wise absolute differences between the two reconstructions (right), as mentioned in \cref{section:methodsresp3d} and \cref{section:resultsresp3d}. The visualization shows data from volunteer 1.}
    \label{video:RRMRMOTUS_vs_RRIR}
\end{figure}

\begin{figure}[h!]
    \centering
    \caption{{\it This is an animated figure and should be viewed at \url{https://surfdrive.surf.nl/files/index.php/s/XcXY3cHETnmXtY0}.} Respiratory-resolved image reconstruction (Resp. resolved IR, left), MR-MOTUS warped reference images (middle), and pixel-wise absolute differences between the two reconstructions (right), as mentioned in \cref{section:methodsresp3d} and \cref{section:resultsresp3d}. The visualization shows data from volunteer 2.}
    \label{video:RRMRMOTUS_vs_RRIR_VOL2}
\end{figure}

\begin{figure}[h!]
    \centering        
    \caption{{\it This is an animated figure and should be viewed at \url{https://surfdrive.surf.nl/files/index.php/s/XcXY3cHETnmXtY0}.} MR-MOTUS warped reference images overlayed with reconstructed dynamic motion-fields from respiratory-sorted data, as mentioned in \cref{section:methodsresp3d} and \cref{section:resultsresp3d}. The image shows a decomposition in the reconstructed components $\bm{\Phi}^i$ (spatial) and $\bm{\Psi}$ (temporal) for volunteer 1. For visualization purposes the components were scaled such that $\lVert \bm{\Phi}^i \rVert=1$.} 
    \label{video:RR_RankDecomp}
\end{figure}

\begin{figure}[h!]
    \centering        
    \caption{{\it This is an animated figure and should be viewed at \url{https://surfdrive.surf.nl/files/index.php/s/XcXY3cHETnmXtY0}.} MR-MOTUS warped reference images overlayed with reconstructed dynamic motion-fields from respiratory-sorted data, as mentioned in \cref{section:methodsresp3d} and \cref{section:resultsresp3d}. The image shows a decomposition in the reconstructed components $\bm{\Phi}^i$ (spatial) and $\bm{\Psi}$ (temporal) for volunteer 2. For visualization purposes the components were scaled such that $\lVert \bm{\Phi}^i \rVert=1$.} 
    \label{video:RR_RankDecomp_VOL2}
\end{figure}

\begin{figure}[h!]
    \centering        
    \caption{{\it This is an animated figure and should be viewed at \url{https://surfdrive.surf.nl/files/index.php/s/XcXY3cHETnmXtY0}.} MR-MOTUS warped reference images overlayed with reconstructed dynamic motion-fields from respiratory-sorted data, as mentioned in \cref{section:methodsresp3d} and \cref{section:resultsresp3d}. The image shows a decomposition in the reconstructed components $\bm{\Phi}^i$ (spatial) and $\bm{\Psi}$ (temporal) for volunteer 1. For visualization purposes the components were scaled such that $\lVert \bm{\Phi}^i \rVert=1$.} 
    \label{video:TR3Dt_Resp_MRMOTUS}
\end{figure}

\begin{figure}[h!]
    \centering        
    \caption{{\it This is an animated figure and should be viewed at \url{https://surfdrive.surf.nl/files/index.php/s/XcXY3cHETnmXtY0}.} MR-MOTUS warped reference images overlayed with reconstructed dynamic motion-fields from respiratory-sorted data, as mentioned in \cref{section:methodsresp3d} and \cref{section:resultsresp3d}. The image shows a decomposition in the reconstructed components $\bm{\Phi}^i$ (spatial) and $\bm{\Psi}$ (temporal) for volunteer 2. For visualization purposes the components were scaled such that $\lVert \bm{\Phi}^i \rVert=1$.} 
    \label{video:TR3Dt_Resp_MRMOTUS_VOL2}
\end{figure}

\begin{figure}[h!]
    \centering        
    \caption{{\it This is an animated figure and should be viewed at \url{https://surfdrive.surf.nl/files/index.php/s/XcXY3cHETnmXtY0}.} MR-MOTUS warped reference images resulting from the 3D head-and-neck motion reconstructions for volunteer 1, as mentioned in \cref{section:methodshead3d} and \cref{section:resultshn3d}.} 
    \label{video:TR3Dt_HN_MRMOTUS}
\end{figure}

\begin{figure}[h!]
    \centering        
    \caption{{\it This is an animated figure and should be viewed at \url{https://surfdrive.surf.nl/files/index.php/s/XcXY3cHETnmXtY0}.} MR-MOTUS warped reference images resulting from the 3D head-and-neck motion reconstructions for volunteer 2, as mentioned in \cref{section:methodshead3d} and \cref{section:resultshn3d}.} 
    \label{video:TR3Dt_HN_MRMOTUS_VOL2}
\end{figure}

\setsuppvideo
\begin{figure}[h!]
    \centering
    \caption{{\it This is an animated figure and should be viewed at \url{https://surfdrive.surf.nl/files/index.php/s/XcXY3cHETnmXtY0}.} Respiratory-resolved image reconstruction (Resp. resolved IR, left), MR-MOTUS warped reference images (middle), and pixel-wise absolute differences between the two reconstructions (right), as mentioned in Supporting Information Section 3 and \cref{section:methodsresp3d}. The four blocks show reconstructions with different reconstruction parameter settings. `InhaleBinned' denotes whether the reference image is binned in inhale (1) or exhale (0). `Ref. resolution' denotes the resolution of the reference image in millimeters. All motion-fields were reconstructed without regularization and with 9 cubic spline functions in every direction.}
    \label{video:RRMRMOTUS_vs_RRIR_4x}
\end{figure}

\setsuppfig
\begin{figure}[h!]
    \centering
    \includegraphics[width=1\textwidth]{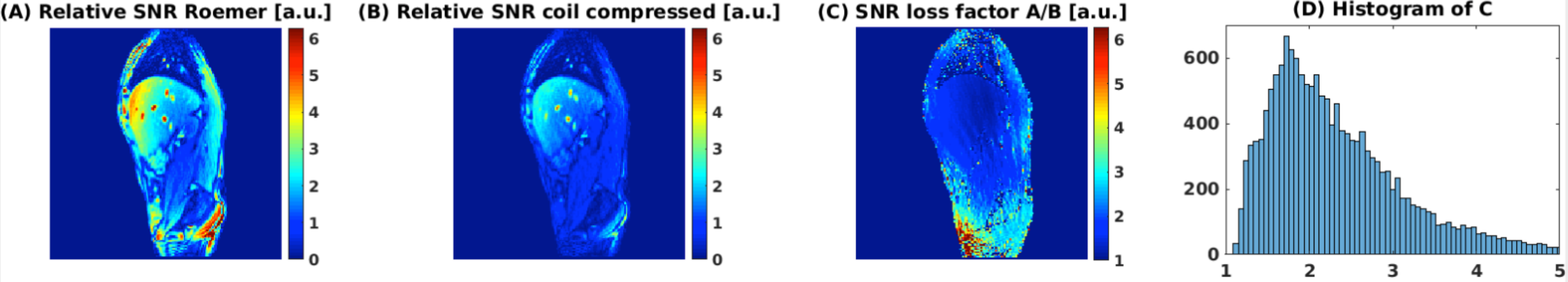}
    \caption{(A) Roemer reconstruction \citep{Roemer1990}. (B) The coil compression with $\lambda_{CC}=5\cdot 10^5$, as discussed in Supporting Information Section 1. (C) SNR loss factor between (A) and  (B). (D) The histogram of the SNR loss factor in (C). The SNR loss factor is between 1.5 and 2.5 in most of the body and increases towards the boundary of the body.}
    \label{fig:2D_snr_comparison}
\end{figure}

\begin{figure}[h!]
    \centering
    \includegraphics[width=\textwidth]{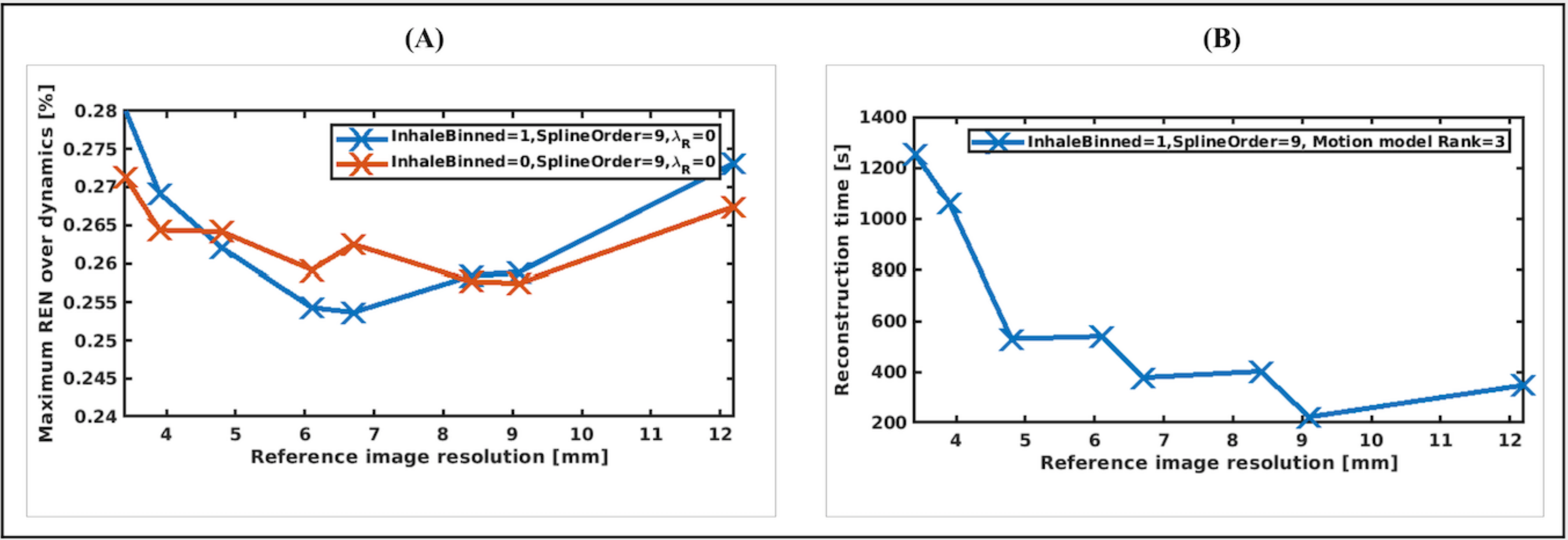}
    \caption{Results of the parameter search as mentioned in Supporting Information Section 3 and \cref{section:methodsresp3d}. (A) The effect of the reference image resolution and reference image respiratory binning-phase on the reconstruction quality. (B) The effect of the reference image resolution on the reconstruction time. In the figure, `InhaleBinned' refers to the binning phase for the reference image (InhaleBinned=1 for inhale, InhaleBinned=0 for exhale), `Resolution' denotes the spatial resolution of the reference image, and `SplineOrder' denotes the number of spline basis functions defined per spatial dimension. }
    \label{fig:RR3DtvsRRIR:errorvsresolution}
\end{figure}

\end{document}